\documentclass[11pt]{article} \usepackage{amsmath, amsthm, amssymb,
  dsfont, mathrsfs}
\usepackage[pdftex,left=1in,top=1in,bottom=1in,right=1in]{geometry}
\usepackage{filecontents} \immediate\write18{bibtex \jobname}
\usepackage[mathscr]{euscript}
\usepackage{graphicx}
\usepackage[initials,alphabetic]{amsrefs}

\title{Expressions for the Entropy of Binomial-Type Distributions\thanks{%
A preliminary version %
of this work appears in Proceedings of the
2018 IEEE International Symposium on Information Theory (ISIT).
}} \author{Mahdi}
\date{}

\newcommand{\cL}{\mathcal{L}}

\newcommand{\E}{\mathds{E}}

\newtheorem{thm}{Theorem} 
 
  \newtheorem{coro}[thm]{Corollary} \theoremstyle{definition}
 
\newtheorem{remark}[thm]{Remark}

\newcommand{\poi}{\mathsf{Poi}}
\newcommand{\hpoi}{H_{\mathsf{Poi}}}
\newcommand{\lam}{\lambda}
\newcommand{\li}{\mathrm{Li}}
\newcommand{\epoi}{E_{\mathsf{Poi}}}
\newcommand{\hepoi}{\hat{E}_{\mathsf{Poi}}}
\newcommand{\bin}{\mathsf{Bin}}

\newcommand{\ebin}{E_{\mathsf{Bin}}}
\newcommand{\fall}[2]{({#1})_{{#2}}}
\newcommand{\rise}[2]{{#1}^{{(#2)}}}
\newcommand{\bbin}{\mathsf{BBin}}

\newcommand{\ebbin}{E_{\mathsf{BBin}}}
\newcommand{\hyper}{{{}_{2}F_{1}}}

\newcommand{\nbin}{\mathsf{NBin}}
\newcommand{\enbin}{E_{\mathsf{NBin}}}
\newcommand{\egeom}{E_{\mathsf{Geom}}}
\newcommand{\hg}{\mathsf{HG}}

\usepackage[hidelinks]{hyperref}

\author{{\sc Mahdi Cheraghchi}\thanks{%
Email: $\langle$m.cheraghchi@imperial.ac.uk$\rangle$. }\\
Department of Computing \\
Imperial College London \\ London, UK 
}

\begin{document}

\maketitle

\begin{abstract}
We develop a general method for computing logarithmic
and log-gamma
expectations of distributions. As a result, we derive
series expansions and integral representations of the 
entropy for several fundamental distributions, including
the Poisson, binomial, beta-binomial, negative
binomial, and hypergeometric distributions. Our results also establish
connections between the entropy functions and to
the Riemann zeta function and its generalizations.
\end{abstract}

\section{Introduction}
\label{sec:intro}
Deriving expressions for the Shannon entropy of commonly studied distributions
is of fundamental significance to information and communication theory, statistics,
and theoretical computer science. For many distributions,
exact closed-form expression for the
entropy is known. A non-comprehensive list of such distributions include
uniform, Bernoulli, geometric, exponential, Laplace, normal, 
log-normal, 
Pareto, Cauchy, Weibull, Rayleigh, $t$-distribution, Dirichlet,
Wishart, Chi-squared, scaled inverse chi-squared, gamma, and,
inverse-gamma distribution. In many cases, the entropy is 
a simple expression in terms of the first few moments, and in
other cases, the logarithmic expectation of the distribution
takes a tractable form. For many fundamental distributions, however,
we do not expect to have direct, closed form, expressions for the entropy
in terms of elementary, or common special, functions. 
In such cases, high quality approximations,
series expansions, or integral forms, for the entropy is desirable. 

In this work, we focus on what we call ``binomial-type'' distributions,
that exhibit factorial terms in their expressions for the probability
mass function. 
In fact, we derive a general expression for
computing log-gamma expectations; i.e., expressions of the form
$\E[\log \Gamma(\alpha+X)]$ for \emph{any} distribution in terms
of its moment generating function. Specific examples that we will
use to demonstrate our technique includes Poisson, binomial,
beta-binomial, negative binomial, and hypergeometric distributions. 
We recall that the binomial
distribution is defined to capture the number of success events
in a series of $n$ Bernoulli trials, for a given parameter $n$ and
a given success probability. This contains the Poisson distribution
as a limiting case, and is in turn a limiting case for the 
more general beta-binomial distribution. A negative binomial
distribution is defined similar to the binomial distribution, 
but with a varying number of trials and fixed number $r$ of success events.
Finally, a hypergeometric distribution is regarded as an analogue
of the binomial distribution where the Bernoulli trials
are performed without replacement.
Apart from their fundamental nature, the entropy
of such distributions
are of particular significance to information theory,
for example in quantifying the number of bits required
for compression of the corresponding sources.
Moreover, understanding the entropy of binomial 
and Poisson distributions is a key step towards a
characterization of the capacity of basic channels with 
synchronization errors, including the deletion and
Poisson-repeat channels (cf.~\cite{ref:Mit08}).
Poisson entropy is also of key significance to the
theory of optical communication and in understanding
the capacity of Poisson-type channels in this context
(cf.~\cite{ref:Sha90,ref:Ver99,ref:Ver08,ref:AW12}).
Indeed, an integral expression for the log-gamma
expectation of Poisson random variables (similar
to one of our results) was the key to \cite{ref:Mar07,ref:CR18}
for obtaining sharp elementary estimates on the capacity of
the discrete-time Poisson channel. On the other hand,
similar integral representations, of the type we derive, 
were used in recent works of the author \cite{ref:Che17,ref:CR18b}
to derive strong upper bounds on the capacity of the binary
deletion, Poisson-repeat, and related channels with synchronization errors.

For a wide range of the parameters, binomial-type distributions are
approximated by a normal distribution via the central limit theorem.
When the variance of these distributions is large (and, for the
negative binomial distribution, the order parameter $r$ is large),  
the entropy is quite accurately estimated by the entropy of a 
normal distribution with matching variance. However, as the variance
decreases, the quality of this approximation deteriorates. 
In fact, as the variance $\sigma$ tends to zero, the entropy of a normal
distribution, which is $\frac{1}{2}\log(2\pi e \sigma^2)$,
diverges to $-\infty$, while the actual entropy of the 
underlying distribution tends to zero. 
In such cases, a more refined expression
for the entropy in terms of a convergent power series or
integral expression would correctly
capture its behavior.

Our starting point is a simple, but curious, manipulation of the
entropy expression for the Poisson distribution. Recall that
a Poisson distribution is defined 
by the probability mass function
\[
\poi(k; \lam) = \frac{\lam^k e^{-\lam}}{k!},
\]
where $k \in \{0,1,\ldots\}$, and the parameter $\lam > 0$
is the mean and variance of the distribution. The entropy of 
this distribution is thus equal, by the definition of Shannon entropy, to
\begin{equation} \label{eqn:Hpoi}
\hpoi(\lam) := -\sum_{k=0}^\infty \poi(k;\lam) \log(\poi(k;\lam))=
-\lam \log(\lam/e) + e^{-\lam} \sum_{k=0}^\infty \frac{\lam^k \log k!}{k!}.
\end{equation}
As mentioned above, $\hpoi(\lam)$ tends to 
$\frac{1}{2}\log(2\pi e \sigma^2)$ as $\lam$ grows. 
A more accurate estimate is given in 
\cite{ref:EB88}; namely, that for large $\lam$,
\[
\hpoi(\lam) = \frac{1}{2}\log(2\pi e \lam) - \frac{1}{12\lam}
-\frac{1}{24 \lam^2}-\frac{19}{360 \lam^3} +O(\lam^{-4}).
\]

We remark that asymptotic expansions for the entropy of binomial
and negative binomial distributions for large mean (in terms of the
difference between the entropy and the Gaussian entropy estimate)
has been derived in \cite{ref:JS99} and
\cite{ref:CGKK13} (and, among other results, in \cite{ref:DV98} and \cite{ref:Kne98}).

Let $X$ be a Poisson distributed random variable with mean $\lam$,
so that the second term on the right hand side of \eqref{eqn:Hpoi}
becomes $\epoi(\lam):=\E[\log X!]$. We now write, using
the convolution formula,
\begin{align}
\epoi(\lam) &= e^{-\lam} \sum_{k=0}^\infty \frac{\lam^k \log k!}{k!} \nonumber \\
&= \sum_{j=0}^\infty \lam^j \sum_{k=0}^j \frac{(-1)^{j-k}}{(j-k)!} \frac{\log k!}{k!}  \nonumber  \\
&= \sum_{j=0}^\infty \frac{\lam^j}{j!} \sum_{k=0}^j \binom{j}{k} (-1)^{j-k} \log k!  \nonumber \\
&= \sum_{j=0}^\infty \frac{(-\lam)^j}{j!} c(j), \label{eqn:epoi}
\end{align}
where we have defined the coefficients
$c(j):=\sum_{k=0}^j \binom{j}{k} (-1)^{k} \log k!$.
The first few values for the $c(j)$, $j=0,1,\ldots$, are
(see integer sequences A122214 and A122215)
\begin{gather*}
\log\left\{1,1,2,\frac{4}{3},\frac{32}{27},\frac{4096}{3645},\frac{67108864}{61509375},\frac{4503599627370496}{4204742431640625},\frac{2535301200456458802993406410752}{2396825584582984447479248046875} \right\} \\
\approx \{0,0,0.693147,0.287682,0.169899,0.116655,0.0871265,0.068664,0.0561673\},
\end{gather*}
where the logarithms are taken to base $e$. 
We have thus derived the Maclaurin series
expansion of the function $\epoi(\lam)$, which converges for all $\lam>0$.
If we formally replace each coefficient $c(j)$ in \eqref{eqn:epoi}
with $(e^{t}-1)^j$, 
the expression turns into the power series expansion of
$e^{\lam(e^{t}-1)}$, which is the moment generating function
of the Poisson distribution. Our main observation is that
this is not a coincidence, and in fact the same phenomenon occurs
for the log-gamma expectation of \emph{any} (discrete or continuous) distribution 
defined over the non-negative reals. In particular, we prove the
following:
\begin{thm} \label{thm:logG}
Let $M(t)$ be the moment generating function of any (continuous
or discrete) distribution with mean $\mu=M'(0)$, and $\alpha > 0$ be a parameter.
Suppose $M(t)$ is analytic around $t=0$ and 
let $Q(z) := M(\log(z+1))$ be represented by power series
$Q(z) = 1+\sum_{j=1}^\infty q(j) z^j$. Then, for a random variable
$X$ sampled from the distribution given by $M$, we have the following:
\begin{align} 
\E[\log\Gamma(X+\alpha)] &= \log\Gamma(\alpha)+ 
\sum_{j=1}^\infty (-1)^j q(j) c_\alpha(j) \label{eqn:ElogGa}\\
&= \log \Gamma(\alpha)+
\int_0^\infty \left( \frac{\mu e^{-t}}{t}-\frac{e^{-\alpha t}(1-M(-t))}{t(1-e^{-t})}\right)dt \label{eqn:ElogGb}\\
&=
\log \Gamma(\alpha)+\mu \log \alpha -
\int_0^1 
\frac{(1-z)^{\alpha-1}}{z \log(1-z)}\, (Q(-z) +\mu z-1) dz, 
\label{eqn:ElogGc}\\
\E[\log(X+\alpha)] &=  
\sum_{j=1}^\infty (-1)^j q(j-1) c_\alpha(j) \label{eqn:ElogGd}\\
&= \int_0^\infty \frac{e^{-t}-e^{-\alpha t} M(-t)}{t}\, dt, \label{eqn:ElogGe}
\end{align}
where 
\begin{align} \label{eqn:coeff}
c_\alpha(j) := -\sum_{k=0}^{j-1} (-1)^{k} \binom{j-1}{k} \log(k+\alpha).
\end{align}
\end{thm}
Proof of the above theorem turns out to be remarkably simple,
yet it provides a general and powerful tool for deriving series expansions
and integral expressions for the entropy of distributions
involving factorial terms in their probability mass functions,
a task that may seem elusive by a direct approach.

\subsection{Summary of the main results}

We apply Theorem~\ref{thm:logG} to derive series and integral expressions
for the entropy and log-gamma expectations of several distributions
over the non-negative integers. We will particularly consider the Poisson,
binomial, beta-binomial, negative binomial, and hypergeometric distributions that are
respectively defined by the probability mass functions below\footnote{
We refer the reader to standard textbooks on probability (such as
\cite{ref:DS12}) and information theory (such as \cite{ref:cover})
for the standard definitions of the various notions used.
} (for
$k=0,1,\ldots$):

\newcommand{\xpoi}{X_{\mathsf{Poi}}}
\newcommand{\xbin}{X_{\mathsf{Bin}}}
\newcommand{\xbbin}{X_{\mathsf{BBin}}}
\newcommand{\xnbin}{X_{\mathsf{NBin}}}
\newcommand{\xgeom}{X_{\mathsf{Geom}}}
\newcommand{\xhg}{X_{\mathsf{HG}}}

\begin{align}
\poi(k; \lam) &= \frac{\lam^k e^{-\lam}}{k!},& \lam>0\\
\label{eqn:bin}
\bin(k; n,p) &= \binom{n}{k} p^k (1-p)^{n-k},& n>0,\ p\in(0,1) \\
\label{eqn:bbin}
\bbin(k;n,\alpha,\beta) &= \binom{n}{k} \frac{B(k+\alpha,n-k+\beta)}{B(\alpha, \beta)},& n,\alpha,\beta>0  \\
\label{eqn:nbin}
\nbin(k; r,p) &= \binom{k+r-1}{k} p^k (1-p)^{r}, & r>0,\ p\in(0,1)\\
\label{eqn:hg}
\hg(k; N,K,n) &= \frac{\binom{K}{k} \binom{N-K}{n-k}}{\binom{N}{n}}, & 
N,K,n > 0
\end{align}
where $B(\alpha,\beta)=\Gamma(\alpha)\Gamma(\beta)/\Gamma(\alpha+\beta)$ 
denotes the beta function. 
We remark that a similar technique has been employed in \cite{ref:Kne98}
(and rediscovered in \cite{ref:Mar07})
to derive integral expressions for the entropy of Poisson, binomial, and
negative binomial distributions.

Let $\xpoi$, $\xbin$, $\xbbin$, 
$\xbin$, and $\xhg$ 
be random variables drawn from their respective distributions above.
In Sections 
\ref{sec:poi},~\ref{sec:bin},~\ref{sec:bbin}, \ref{sec:nbin},
and \ref{sec:hg},
we derive the expressions below.
All expressions converge for the whole range of parameters,
and logarithms are taken to the base in which the entropy is measured.

\begin{align}
\label{eqn:hpoi}
H(\xpoi)&=-\lam \log(\lam/e)+\sum_{j=2}^\infty \frac{c(j)}{j!}\, (-\lam)^j
 \\
&=-\lam \log(\lam/e)+\int_0^1 
\frac{1-e^{-\lam z} -\lam z}{z \log(1-z)}\, dz, \label{eqn:hpoiInt}
\end{align}
where we use the convention\footnote{We note that 
the expression for $c(j)$ in \eqref{eqn:epoi} indeed coincides with $c_1(j)$
due to the fact that the logarithm function is the first finite
derivative of log-gamma.
} $c(j) := c_1(j)$.

\begin{align} \label{eqn:hbinIntSummary}
H(\xbin) &= n h(p) + \int_0^\infty \left( \frac{(1-p+p e^{-t})^n+
(p+(1-p) e^{-t})^n-e^{-nt}-1}{t(e^t-1)}\right)dt \\
&= n h(p) + \sum_{j=2}^\infty \binom{n}{j} (-1)^j c(j) (p^j+(1-p)^j-1),
\label{eqn:hbinSummary}
\end{align}
 where $h(p) := -p \log p -(1-p) \log(1-p)$ is the binary entropy
 function.
\begin{align}
H(\xbbin)&= 
-\log (n B(n,\alpha+\beta))
+\sum_{j=2}^\infty \binom{n}{j} \frac{(-1)^j}{\rise{(\alpha+\beta)}{j}}
((c(j)-c_\alpha(j)) \rise{\alpha}{j} + 
(c(j)-c_\beta(j)) \rise{\beta}{j}), \label{eqn:HbbinSeriesSummary}
\end{align}
where we use the notation $\rise{a}{j}$ for the rising factorial
$a(a+1)\cdots(a+j-1)$.
\begin{align}
\E[\log \Gamma(\xbbin+r)] \label{eqn:EbbinIntSummary}
&= \log\Gamma(r)+\frac{n \alpha \log r}{\alpha+\beta}\\&+
\int_0^1 \frac{(1-z)^{r-1} (\hyper(-n,\alpha;\alpha+\beta;-z)+
n \alpha/(\alpha+\beta)-1)}{z \log(1-z)}\,dz \\
&= \log\Gamma(r)+
\sum_{j=2}^\infty (-1)^j c_r(j) \frac{\rise{\alpha}{j}}{\rise{(\alpha+\beta)}{j}} \binom{n}{j}.
\label{eqn:EbbinSumSummary}
\end{align}

\begin{align} \label{eqn:hnbinSummary}
H(\xnbin)&=\frac{r h(p)}{1-p}+
\sum_{j=1}^\infty \binom{j+r-1}{j} \left(\frac{p}{p-1}\right)^j (c(j)-c_r(j)) \\
&=\frac{r(h(p)-p \log r)}{1-p} +
\int_0^1 
\frac{((1-z)^{r-1}-1)((1+pz/(1-p))^{-r} +pr z/(1-p)-1)}{z \log(1-z)}\,  dz.
\label{eqn:hnbinIntSummary}
\end{align}

\begin{align} 
H(\xhg)&=\log \binom{N}{n}-\log K!-\log(N-K)! \nonumber \\&+
\sum_{j=2}^\infty \frac{(-1)^j c(j)}{\fall{N}{j}} \left(
\fall{K}{j} \fall{n}{j} +
\fall{N-K}{j} \fall{n}{j} +
\fall{K}{j} \fall{N-n}{j} +
\fall{N-K}{j} \fall{N-n}{j}
\right). \label{eqn:hhgSummary}
\end{align}

We demonstrate examples of connections between the entropy functions 
via functional transformations as well as connections between them
and the Riemann zeta function and its many generalizations. We believe that
such connections will stimulate further research towards a full
understanding of the entropy of such fundamental distributions as
the Poisson, binomial, and related distributions.
 
Among our results, we show (in Section~\ref{sec:poi}, Theorem~\ref{thm:Laplace}) that 
the Laplace transform of $H(\xpoi)$,
regarded as a function of the expectation $\lam$, is equal to
\[
\frac{\gamma+\log z}{z^2}-
\frac{1}{z(1+z)}\, \Phi'\left(\frac{1}{1+z}\right),
\]
where $\Phi'$ is the derivative of the polylogarithm function
as defined in $\eqref{eqn:plog}$. Another example is a connection
between the geometric distribution (on $0,1,\ldots$) and the
Poisson distribution. Letting $\xgeom$ denote a geometrically
distributed random variable, we show (in \eqref{eqn:geomCa}) that 
the logarithmic expectation $\E[\log(\xgeom+1)]$,
as a function of the mean of $\xgeom$, generates the
logarithmic difference coefficients $c_\alpha(j)$ defined in \eqref{eqn:coeff}. 
Moreover, we express this function in terms of the 
Laplace transform of the entropy of a Poisson distribution
(Section~\ref{sec:nbin}, Theorem~\ref{thm:LaplaceG}).
Finally, we derive connections between the $c_\alpha(j)$
and their generating function to the Riemann zeta function
(Appendix~\ref{sec:zeta}) and its generalizations (Section~\ref{sec:cj},
\eqref{eqn:CgenLerch}) that we believe may serve as a natural continuation point
towards a complete understanding of the entropy of Poisson and
related distributions.

\paragraph{Organization.}
The rest of the article is organized as follows. In Section~\ref{sec:cj},
we study the coefficients $c_\alpha(j)$ in \eqref{eqn:coeff} and
their various properties. Section~\ref{sec:proof} gives a proof
of Theorem~\ref{thm:logG}. In Sections 
\ref{sec:poi},~\ref{sec:bin},~\ref{sec:bbin}, \ref{sec:nbin},
and \ref{sec:hg},
we respectively apply Theorem~\ref{thm:logG} to obtain expressions
for the entropy and log-gamma expectations of the Poisson,
binomial, beta-binomial, negative binomial, and hypergeometric distributions.
We conclude in Section~\ref{sec:discussion} by a brief discussion
of possible future directions and questions raised by this work.

\section{The logarithmic difference coefficients $c_\alpha(j)$}
\label{sec:cj}

\begin{figure}[t!]
\begin{center}
Acknowledgement\includegraphics[height=3in]{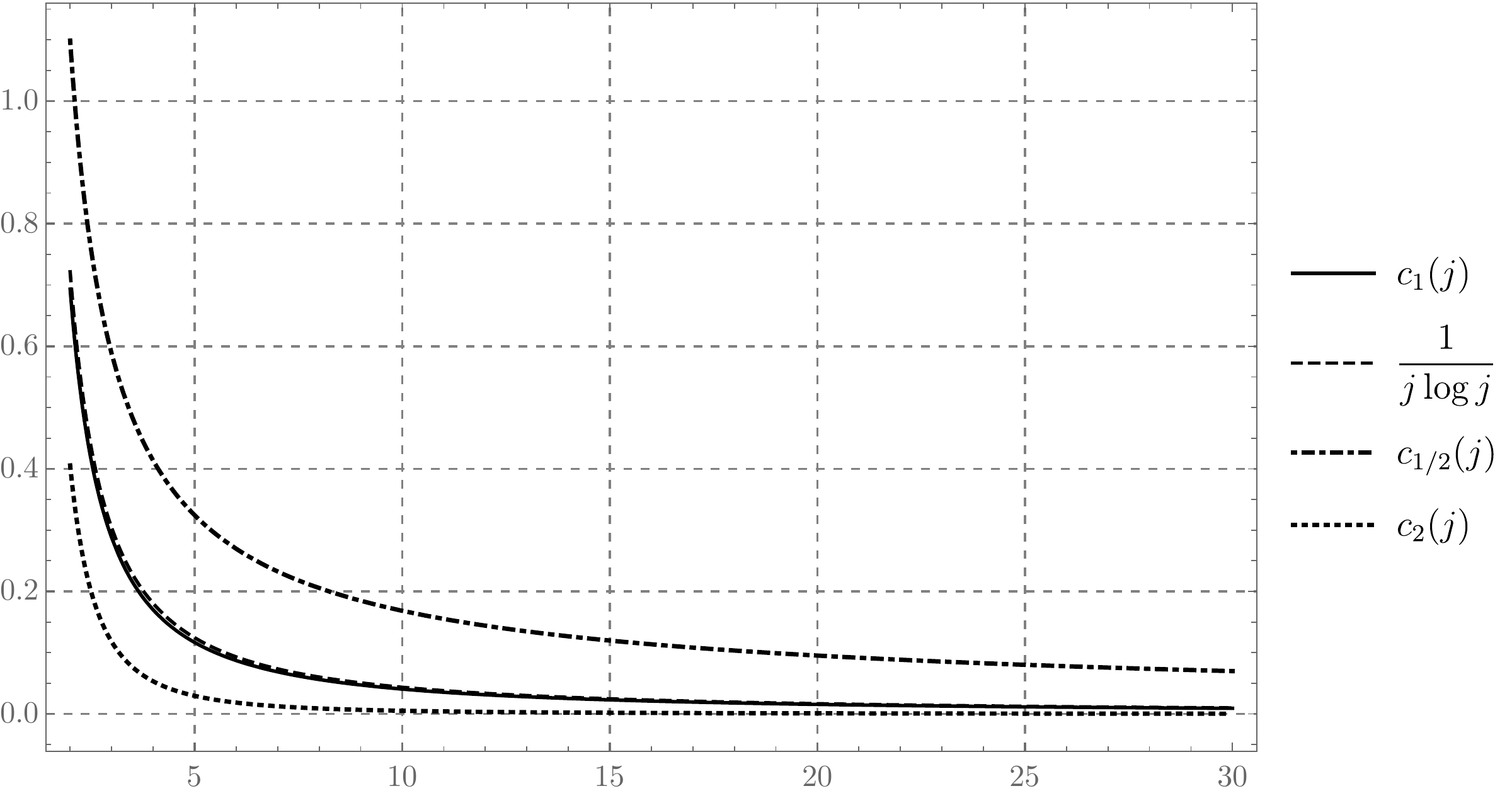}
\end{center}
\caption{Logarithmic difference coefficients $c_\alpha(j)$ as a function
of $j$.}
\label{fig:coeff}
\end{figure}

The coefficients $c_\alpha(j)$ defined in \eqref{eqn:coeff}
naturally appear in the analytic study of zeta functions (we will
study an example related to this work in Appendix~\ref{sec:zeta}) and 
are of fundamental importance in calculus of finite differences. 
They are essentially the Newton series expansion 
coefficients of the logarithm function
around point $\alpha$, which is why we call them the logarithmic
difference coefficients.
The coefficients are plotted\footnote{The plot
depicts the continuous interpolation of the $c_\alpha(j)$ given by
\eqref{eqn:intCoeff}, so that the values are meaningful for
non-integer choices of $j$ as well.
} for various choices of $\alpha$
in Figure~\ref{fig:coeff}. We observe that the function $1/(j \log j)$
closely approximates $c_1(j)$.

\noindent Recall that the $j$th forward difference of
a function $f$ at point $\alpha$ is defined as
\begin{equation} \label{eqn:delta}
\Delta^{j}[f](\alpha) := \sum_{k=0}^j \binom{j}{k} (-1)^{j-k} f(k+\alpha),
\end{equation}
or equivalently, as the $j$-fold application of the discrete
derivative $\Delta[f](\alpha) := f(\alpha+1)-f(\alpha)$ of the
function at $\alpha$ (and letting $\Delta^0[f](\alpha)=f(\alpha)$. 
Furthermore, we recall that the Newton series
expansion of a function $f$ around point $\alpha$ is given by
\begin{equation} \label{eqn:Newton}
f(x+\alpha)=\sum_{j=0}^\infty \binom{x}{j} \Delta^{j}[f](\alpha),
\end{equation}
which is the same as the formula for the Taylor expansion with 
continuous derivatives replaced by forward differences and powers of $x$
replaced by factorial powers. Note that $x$ need not be an integer.
Applying \eqref{eqn:delta} on $f(x)=\log x$, and comparing
with \eqref{eqn:coeff}, we see that
\[
\Delta^{j}[\log](\alpha) = (-1)^{j+1} c_\alpha(j+1),
\]
so that
\[
\log(x+\alpha)=\sum_{j=0}^\infty \binom{x}{j} (-1)^{j+1} c_\alpha(j+1).
\]
Unlike the Taylor expansion, one can verify that the above 
series converges to $\log(x+\alpha)$
for all values of $x > -\alpha$.

The coefficients $c_\alpha(j)$ have a compact integral representation.
When $j>1$, we have
\begin{align} \label{eqn:intCoeff}
c_\alpha(j)=\int_0^\infty \frac{(1-e^{-t})^{j-1} e^{- \alpha t}}{t}\, dt,
\end{align}
which can be readily verified by a binomial expansion of the integrand
and using the following basic identity (which holds for any $n>0$) 
on each resulting term:
\begin{equation} \label{eqn:logn}
\int_{0}^\infty \frac{e^{-t}-e^{-n t}}{t}\, dt=\log n.
\end{equation}

It is worthwhile to understand the generating function
for the coefficients
$c_\alpha(j)$. In order to do so, we start by recalling 
Lerch transcendent (cf.~\cite[p.~27]{ref:EMOT53})
\begin{equation} \label{eqn:Lerch}
\Phi (z,s,\alpha ):=\sum _{k=0}^{\infty }{\frac {z^{k}}{(k+\alpha )^{s}}}
= {\frac  {1}{\Gamma (s)}}\int _{0}^{\infty }{\frac  {t^{{s-1}}e^{{-\alpha t}}}{1-ze^{{-t}}}}\,dt.
\end{equation}
By taking derivative of the above series with respect to $s$ at $s=0$,
we obtain the generating function of the logarithmic sequence
\begin{equation} \label{eqn:lerch}
\Phi'_\alpha(z) := \left.\frac{d}{ds} \Phi(z,s,\alpha)\right|_{s=0}=
-\sum_{k=0}^\infty \log(k+\alpha) z^k.
\end{equation}
When $\alpha=1$, in which case we drop the subscript from the notation,
the above can be written in terms of
the polylogarithm function $\li_s(z):=\sum_{k=1}^\infty z^k k^{-s}$ 
as
\begin{equation}
\label{eqn:plog}
\Phi'(z) := \Phi'_1(z)=\left.\frac{d}{ds} \li_s(z)/z\right|_{s=0}.
\end{equation}

Now, observe that the coefficient sequence $(-c_\alpha(j+1))_{j=0}^\infty$
is the binomial transform \cite[p.~136]{ref:art3} of the sequence $(\log(k+\alpha))_{k=0}^\infty$.
Let 
\begin{equation}
\label{eqn:Cgen}
C_\alpha(z) := \sum_{j=0}^\infty c_\alpha(j) z^j
\end{equation}
be the generating function for $(c_\alpha(j))_{j=0}^\infty$,
where we define $c_\alpha(0):=0$.
Although we may treat the series formally, note that \eqref{eqn:Cgen}
converges when $|z| < 1$ or $z=-1$ (since $c_\alpha(j)$ is a decreasing sequence
for $\alpha>0$).
Using \eqref{eqn:Lerch} and the formula for generating function of the binomial transform,
we thus have
\begin{equation} \label{eqn:CgenLerch}
C_\alpha(z) = \frac{z}{1-z} \Phi'_\alpha\left(\frac{-z}{1-z}\right).
\end{equation}
In this regard, \eqref{eqn:ElogGa} in Theorem~\ref{thm:logG} can be
interpreted as the assertion that the log-gamma expectation 
$\E[\log \Gamma(X+\alpha)]-\log \Gamma(\alpha)$ 
of any distribution over non-negative reals
is the inner product of the power series coefficients 
of $C_\alpha(-z)$
and that of the function $Q(z) := M(\log(z+1))$ derived from the moment generating
function (the power series coefficients of $Q(z)$, 
in turn, are the factorial moments of the distribution).
In other words, the Hadamard product of the functions $C_\alpha(-z)$
and $Q(z)$ evaluated at $z=1$ is equal to 
$\E[\log \Gamma(X+\alpha)]-\log \Gamma(\alpha)$.

We demonstrate another characterization of the generating
function $C_\alpha(z)$ in Section~\ref{sec:nbin}. Namely, 
we will show in \eqref{eqn:enbin} that 
$C_\alpha(-z)$ is the log-gamma expectation of a geometric
distribution over $0,1,\ldots$ with mean $z$.
Furthermore, in Appendix~\ref{sec:zeta}, we present
an intriguing connection between the logarithmic difference
coefficients (and their generating function) and the
Riemann zeta function and its generalized form, the Hurwitz zeta
function. This leads to a formula for a weighted summation
of the coefficients $c_\alpha(j)$ in terms of the digamma function and
Harmonic numbers.

\section{Proof of Theorem~\ref{thm:logG}}
\label{sec:proof}

We consider the Newton series expansion of the function
$\log\Gamma(x)$ around point $\alpha$ given by
\eqref{eqn:Newton}, or equivalently, the series expansion
of $f(x) := \log\Gamma(x+\alpha)$ around zero. 
The discrete derivative of $f$
is given by
\[
\Delta[f](x) = f(x+1)-f(x)=\log\frac{\Gamma(x+\alpha+1)}{\Gamma(x+\alpha)}
=\log(x+\alpha).
\]
Therefore, for $j>0$, the 
$j$th forward difference of $f$ is the $(j-1)$st forward
difference of the logarithmic function $g(x) := \log(x+\alpha)$.
This can be written down, via \eqref{eqn:delta}, as
\begin{align}
\Delta^{j}[f](0) &= \Delta^{j-1}[g](0) \nonumber \\ 
&= -\sum_{k=0}^{j-1} \binom{j-1}{k} (-1)^{j-k} \log(k+\alpha) \nonumber \\
&\stackrel{\eqref{eqn:coeff}}{=} (-1)^j c_\alpha(j).\label{eqn:deltaJf} 
\end{align}
Recall that the factorial moment generating function of a
distribution with moment generating function $M(t)$ is given by
$M(\log z)$, and the coefficients of the power series expansion
of this function around $z=1$ determine the factorial moments.
Therefore, the power series expansion of the function $Q(z)=M(\log(z+1))$
(which is also understood as generating the inverse \emph{Stirling transform}
of the moment sequence) around $z=0$, namely $Q(z)=\sum_{j=0}^\infty q(j) z^j$, 
determines the factorial moments
of the distribution. The $j$th factorial moment of $X$ sampled
from the distribution defined by $M$ is given by
$\E[\fall{X}{j}]=j! q(j)$, where $\fall{X}{j}$ denotes falling factorial.
Note that $q(0)=1$ and $q(1)=\E[X]=\mu$.
Using \eqref{eqn:deltaJf}, we can write down the (convergent)
Newton expansion of $f(x)$ and take its expectation using the information
$Q(z)$ gives on the factorial moments as follows:
\begin{align}
\E\log\Gamma(X+\alpha)&\stackrel{\eqref{eqn:Newton}}{=} \E\left[\log\Gamma(\alpha)+
\sum_{j=1}^\infty \binom{X}{j} (-1)^j c_\alpha(j)\right] \nonumber \\
&= \log\Gamma(\alpha)+\sum_{j=1}^\infty (-1)^j q(j) c_\alpha(j).\nonumber 
\end{align}
This proves \eqref{eqn:ElogGa}. In order to derive \eqref{eqn:ElogGb},
we use \eqref{eqn:intCoeff} in the above result for $j>1$, 
and noting that $c_\alpha(1)=-\log \alpha$, which gives
\[
\E\log\Gamma(X+\alpha)=
\log\Gamma(\alpha)+\mu \log(\alpha)+\sum_{j=2}^\infty (-1)^j q(j) 
\int_0^\infty \frac{(1-e^{-t})^{j-1} e^{- \alpha t}}{t}\, dt.
\]
Recall that $M(-t)=Q(e^{-t}-1)$, and using this in the above expression, 
we may 
change the order of (convergent) summation and integration and write
\begin{align}
\E[\log\Gamma(X+\alpha)]&=
\log\Gamma(\alpha)+\mu \log \alpha +\int_0^\infty \frac{e^{- \alpha t}}{t(1-e^{-t})} 
\sum_{j=2}^\infty (-1)^j q(j) (1-e^{-t})^{j}  dt \nonumber \\
&= \log\Gamma(\alpha)+\mu \log \alpha+\int_0^\infty \frac{e^{- \alpha t}}{t(1-e^{-t})} 
(Q(e^{-t}-1)-1-\mu(e^{-t}-1)) \nonumber  dt\\
&\stackrel{\eqref{eqn:logn}}{=} 
\log\Gamma(\alpha)+
\int_{0}^\infty \left(\frac{\mu e^{-t}- \mu e^{-\alpha t}}{t}+
\frac{e^{- \alpha t}(M(-t)-1)}{t(1-e^{-t})} 
+\frac{\mu e^{-\alpha t}}{t}\right) \nonumber  dt\\ 
&= 
\log\Gamma(\alpha)+
\int_{0}^\infty \left(\frac{\mu e^{-t}}{t}-
\frac{e^{- \alpha t} (1-M(-t))}{t(1-e^{-t})}\right)  dt \nonumber,
\end{align}
which proves \eqref{eqn:ElogGb}. Now, \eqref{eqn:ElogGc}
can be simply verified by rewriting the above
integral expression in terms of the variable $z=e^{-t}-1$.
Let $\overline{Q}(z) := Q(z) -\mu z-1=\sum_{j=2}^\infty q(j) z^j$.
We have
$t=-\log(1+z)$ and $dt=-dz/(1+z)=-e^t dz$ so that \eqref{eqn:ElogGb}
becomes
\begin{align}
\E[\log \Gamma(X+\alpha)]&= \log \Gamma(\alpha)+
\int_0^\infty \left( \frac{\mu e^{-t}}{t}-
\frac{\mu e^{-\alpha t}}{t}\right)dt-
\int_0^1 
\frac{(1-z)^{\alpha-1}}{z \log(1-z)}\, \overline{Q}(-z) dz \nonumber \\
&\stackrel{\eqref{eqn:logn}}{=}
\log \Gamma(\alpha)+\mu \log \alpha -
\int_0^1 
\frac{(1-z)^{\alpha-1}}{z \log(1-z)}\, \overline{Q}(-z) dz, \nonumber
\end{align}
which proves \eqref{eqn:ElogGc}.
In fact, applying the same change of variables to \eqref{eqn:intCoeff}
shows that, for $j > 1$,
\begin{align} \nonumber
c_\alpha(j) = -\int_0^1 
\frac{(1-z)^{\alpha-1} z^{j-1}}{\log(1-z)}\, dz.
\end{align}
In order to derive \eqref{eqn:ElogGd}, we repeat the Newton expansion 
but directly on the logarithmic function $g$, noting,
from \eqref{eqn:deltaJf}, 
that $\Delta^j[g](0)=(-1)^{j+1} c_\alpha(j+1)$. 
So we have
\begin{align}
\E\log(X+\alpha)&\stackrel{\eqref{eqn:Newton}}{=} \E\left[
\sum_{j=0}^\infty \binom{X}{j} (-1)^{j+1} c_\alpha(j+1)\right] \nonumber \\
&= \sum_{j=1}^\infty (-1)^j q(j-1) c_\alpha(j).\nonumber 
\end{align}
Finally, \eqref{eqn:ElogGe} immediately follows by writing $\log(X+\alpha)$
in integral form using \eqref{eqn:logn} and taking the
expectation of the integrand.

\section{Entropy of the Poisson distribution}
\label{sec:poi}
As the first application of Theorem~\ref{thm:logG}, 
consider a Poisson distributed random variable $X$ 
with mean $\lam$,
and define $\epoi(\lam):=\E[\log X!]$, so that we have
$\hpoi(\lam):=H(X) = -\lam \log(\lam/e) + \epoi(\lam)$.
The generating function of the distribution is
$M(t)=\exp(\lam (e^t-1))$, so we have 
\[
Q(z)=M(\log(z+1))=e^{\lam z}=\sum_{j=0}^\infty \frac{\lam^j}{j!}\, z^j.
\]
Theorem~\ref{thm:logG}, applied with $\alpha=1$ in \eqref{eqn:ElogGa}, 
now directly implies that
\begin{equation} \label{eqn:epoiB}
\epoi(\lam)=\sum_{j=2}^\infty \frac{c(j)}{j!}\, (-\lam)^j,
\end{equation}
where we recall the shorthand $c(j):=c_1(j)$, 
thus recovering the Maclaurin series expansion of
$\epoi$ in \eqref{eqn:epoi}.
Note that this expansion is the same as the Newton series expansion of the function
$\log(\lam!)$, with factorial powers of $\lam$ replaced by actual powers.	
Since $(c(j))_{j=2}^\infty$ is a
decreasing sequence, a simple ratio test reveals that 
the above power series expansion absolutely converges for all
$\lam>0$.
Furthermore, observe that the integral expression \eqref{eqn:hpoiInt}
can be immediately recovered from \eqref{eqn:ElogGc} in 
Theorem~\ref{thm:logG}.

While we do not know of a representation of the function $\epoi(\lam)$
in terms of elementary or natural special functions, we may see that
its Laplace transform (which we will revisit later in Section~\ref{sec:nbin}) 
takes an interesting form. 
Let $\hepoi(z)$ denote the Laplace transform of $\epoi(\lam)$.
By the Laplace
transform formula for power series (namely, $\cL\{ \lam^j \} = j! (1/z)^{j+1}$), 
and using \eqref{eqn:epoiB}, we may write
\begin{equation} \label{eqn:LaplaceP}
\hepoi(z) = -\sum_{j=2}^\infty c(j) (-1/z)^{j+1}
\stackrel{\eqref{eqn:CgenLerch}}{=} \frac{-1}{z(1+z)}\, \Phi'\left(\frac{1}{1+z}\right).
\end{equation}
Combined with the Laplace transform of $-\lam \log(\lam/e)$,
which is $(\gamma+\log z)/z^2$ (where $\gamma \approx 0.57721$ 
is the Euler-Mascheroni constant), we conclude the following:

\begin{thm} \label{thm:Laplace}
The Laplace transform of the entropy function $\hpoi(\lam)$ is given by
\[
\cL\{\hpoi\}(z) = \frac{\gamma+\log z}{z^2}-
\frac{1}{z(1+z)}\, \Phi'\left(\frac{1}{1+z}\right),
\]
where $\Phi$ is derivative of the polylogarithm function
defined in \eqref{eqn:plog}.
\end{thm}

\begin{remark}
An alternative way of proving Theorem~\ref{thm:Laplace} is
to directly start from \eqref{eqn:Hpoi} and use the
frequency shifting property of the Laplace transform
and then the convolution property of generating functions
(applied on $\Phi'(z)$ and the geometric series $1/(1-z)$)
in order to generate the sequence $(\log k!)_{k=0}^\infty$
from $(\log (1+k))_{k=0}^\infty$.
\end{remark}

\section{Entropy of the binomial distribution}
\label{sec:bin}
Recall that the binomial distribution with parameters
$n, p$ is defined by the probability mass function in \eqref{eqn:bin}.
In general, $n$ need not be an integer.
This distribution has mean $np$, variance $np(1-p)$,
and moment generating function $(1-p+p e^{t})^n$.

Let $X$ be a random variable distributed according to 
a binomial distribution with parameters $n, p$. Similar 
to the Poisson distribution, the difficulty in computing
$H(X)$ is captured by the computation of
\[
\ebin(n,p):=\E[\log X!],
\] 
and, directly from \eqref{eqn:bin},
 we can see that $H(X)=n h(p)-\log\Gamma(n+1)+\ebin(n,p)+\ebin(n,1-p)$.

Using \eqref{eqn:ElogGb}, we can immediately write down an
integral representation of $\ebin(n,p)$:
\begin{align*}
\ebin(n,p) &= \int_0^\infty \left( \frac{n p e^{-t}}{t}-\frac{e^{-t}(1-(1-p+p e^{-t})^n)}{t(1-e^{-t})}\right)dt,
\end{align*}
so that we have the integral representation of $H(X)$ given by \eqref{eqn:hbinIntSummary}:
\begin{align} \nonumber 
H(X) = n h(p) + \int_0^\infty \left( \frac{(1-p+p e^{-t})^n+
(p+(1-p) e^{-t})^n-e^{-nt}-1}{t(e^t-1)}\right)dt.
\end{align}
Now, observe that the function $Q(z)$ in Theorem~\ref{thm:logG}
is $Q(z) = (1-p z)^n = \sum_{j=0}^\infty \binom{n}{j} (-p z)^j $.
Therefore, by \eqref{eqn:ElogGa}, we have
\begin{align} \label{eqn:ebin}
\ebin(n,p)=\sum_{j=2}^\infty \binom{n}{j} c(j) (-p)^j = 
\sum_{j=2}^\infty \frac{c(j)}{j!}\, (-p)^j \fall{n}{j},
\end{align}
where $\fall{n}{j}$ denotes the falling factorial.
This gives us the following series expansion for the entropy:
\begin{equation} \nonumber
H(X)= n h(p) + \sum_{j=2}^\infty \binom{n}{j} (-1)^j c(j) (p^j+(1-p)^j-1),
\end{equation}
thus confirming \eqref{eqn:hbinSummary}.

\begin{remark}
Observe the remarkable similarity between \eqref{eqn:ebin} and
the analogous quantity $\epoi$ for the Poisson distribution in \eqref{eqn:epoi}:
Indeed, \eqref{eqn:ebin} is obtained from \eqref{eqn:epoi} by letting
$\lam=np$ and replacing powers of $n$ with factorial powers of the same
order. This makes intuitive sense as the Poisson distribution is 
the limiting case for the binomial distribution as $n$ tends to infinity
and the expectation $np$ is fixed to the desired parameter $\lambda$,
in which case falling factorials $\fall{n}{j}$ are within a multiplicative
factor $1+O(1/n)$ of the corresponding actual powers $n^j$, a factor that
tends to $1$ as $n$ grows. \qed
\end{remark}


\section{Entropy of the beta-binomial distribution}
\label{sec:bbin}
A beta-binomial distribution is defined by positive parameters
$n, \alpha, \beta$ (where $n$ is typically an integer)
and probability mass function $\bbin(k;n,\alpha,\beta)$
in \eqref{eqn:bbin}.
The moment generating function for this distribution is
\[
M(t) = \hyper(-n,\alpha;\alpha+\beta;1-e^{t}),
\]
where $t < \log 2$ and $\hyper$ denotes the hypergeometric function defined as
\begin{equation} \label{eqn:hyper}
\hyper(a,b;c;z)=\sum _{j=0}^{\infty }{\frac {\rise{a}{j} \rise{b}{j}}{\rise{c}{j}}}{\frac {z^{j}}{j!}},
\end{equation}
and $\rise{a}{j}$ is the rising factorial.
Let $X$ be a random
variable with beta distribution defined by parameters $n, \alpha, \beta$.
The distribution of $X$ is a compound 
 distribution generated by
sampling a random parameter $p \in (0,1)$ according to a beta
distribution with parameters $\alpha, \beta$, and subsequently, 
drawing from a binomial distribution defined by parameters $n, p$.
The expectation of $p$ is $\alpha/(\alpha+\beta)$ (and thus
$\E[X]=n \alpha/(\alpha+\beta)$). For a fixed 
ratio $\alpha/(\alpha+\beta)$, the variance of $p$ decreases as
$\alpha$ grows, and thus the binomial distribution is a limiting case
of the beta-binomial distribution when $\alpha \rightarrow \infty$
and $\alpha/(\alpha+\beta)=p$.

Similar to the Poisson and binomial distributions, the difficulty
in computing the entropy of a beta-binomial distribution lies at
the computation of
\[
\ebbin(n,\alpha,\beta, r) := \E[\log \Gamma(X+r)],
\]
for a fixed parameter $r$ (which is either $1$, $\alpha$, or $\beta$).
Using this notation, and noting that the distribution of
$n-X$ is given by $\bbin(k;n, \beta,\alpha)$, the entropy can be written as
\begin{align}
H(X)&=\log B(\alpha,\beta)+\E \left[\log\frac{\Gamma(n+\alpha+\beta)\Gamma(X+1) \Gamma(n-X+1)}
{\Gamma(n+1) \Gamma(X+\alpha) \Gamma(n-X+\beta)}\right] \nonumber\\
&= \log B(\alpha,\beta)+\log\frac{\Gamma(n+\alpha+\beta)}{\Gamma(n+1)}
+\ebbin(n,\alpha,\beta,1)+\ebbin(n,\beta,\alpha,1) \nonumber\\
&-\ebbin(n,\alpha,\beta,\alpha)-\ebbin(n,\beta,\alpha,\beta). \label{eqn:Hbbin}
\end{align}
As before, we may use \eqref{eqn:ElogGb} and \eqref{eqn:ElogGc},
noting that $Q(z)=\hyper(-n,\alpha;\alpha+\beta;z)$, to derive an integral
representation of the function $\ebbin$ as
\begin{align*}
\ebbin(n,\alpha,\beta, r) &\stackrel{\eqref{eqn:ElogGb}}{=} \log\Gamma(r)+\int_0^\infty \left( \frac{n \alpha e^{-t}}{t(\alpha+\beta)}-\frac{e^{-r t}(1-\hyper(-n,\alpha;\alpha+\beta;1-e^{t}))}{t(1-e^{-t})}\right)dt\\
&\stackrel{\eqref{eqn:ElogGc}}{=} \log\Gamma(r)+\frac{n \alpha \log r}{\alpha+\beta}+
\int_0^1 \frac{(1-z)^{r-1} (\hyper(-n,\alpha;\alpha+\beta;-z)+
n \alpha/(\alpha+\beta)-1)}{z \log(1-z)}\,dz,
\end{align*}
which confirms \eqref{eqn:EbbinIntSummary}.
By \eqref{eqn:hyper}, the coefficient of $z^j$ in the series
expansion of $Q(z)$ is equal to
\[
\frac{(-1)^j \rise{\alpha}{j}}{\rise{(\alpha+\beta)}{j}} \binom{n}{j}.
\]
Therefore, using \eqref{eqn:ElogGa}, we obtain the series expansion 
of $\ebbin(n,\alpha,\beta, r)$ in \eqref{eqn:EbbinSumSummary}:
\begin{equation} \label{eqn:ebbinSeries}
\ebbin(n,\alpha,\beta, r)=\log\Gamma(r)+
\sum_{j=2}^\infty (-1)^j c_r(j) \frac{\rise{\alpha}{j}}{\rise{(\alpha+\beta)}{j}} \binom{n}{j}.
\end{equation}
Observe that, letting $p:=\alpha/(\alpha+\beta)$, if $p$ is fixed while
$\alpha$ grows large, the ratio $\rise{\alpha}{j}/\rise{(\alpha+\beta)}{j}$
becomes $p^j(1+O(1/\alpha))$, and thus, the terms
in \eqref{eqn:ebbinSeries} converge to those in
\eqref{eqn:ebin} (albeit \eqref{eqn:ebin} is written for the special
case $r=1$). This is consistent with the fact that the binomial distribution
is the limiting distribution of the beta-binomial distribution
for fixed $p$ as
$\alpha$ tends to infinity. Plugging this result into 
\eqref{eqn:Hbbin}, we derive a series expansion for the entropy
of beta-binomial distribution:
\begin{align}
H(X)&= \log \left(\frac{\Gamma(\alpha) \Gamma(\beta)}{\Gamma(\alpha+\beta)}
\frac{\Gamma(n+\alpha+\beta)}{n \Gamma(n)}\right)-
\log \Gamma(\alpha)-\log \Gamma(\beta)\nonumber \\
&+\sum_{j=2}^\infty \binom{n}{j} \frac{(-1)^j}{\rise{(\alpha+\beta)}{j}}
((c(j)-c_\alpha(j)) \rise{\alpha}{j} + 
(c(j)-c_\beta(j)) \rise{\beta}{j}) \nonumber \\
&= 
-\log (n B(n,\alpha+\beta))
+\sum_{j=2}^\infty \binom{n}{j} \frac{(-1)^j}{\rise{(\alpha+\beta)}{j}}
((c(j)-c_\alpha(j)) \rise{\alpha}{j} + 
(c(j)-c_\beta(j)) \rise{\beta}{j}), \nonumber 
\end{align}
which proves \eqref{eqn:HbbinSeriesSummary}.

\section{Entropy of the negative binomial distribution}
\label{sec:nbin}
Recall that the negative binomial distribution is defined
by the probability mass function $\nbin(k; r,p)$ in \eqref{eqn:nbin},
for parameters $r>0$ and $p \in (0,1)$, and has mean $p r/(1-p)$.
When $r$ is an integer, the distribution captures an independent summation of
$r$ identical, geometrically distributed, random variables.
The moment generating function of the distribution is given by
\[
M(t) = \left(\frac{1-p}{1-p e^t}\right)^r.
\]
Let $X$ be a negative binomial random variable with parameters
$r$ and $p$, and define $\ebin(r,p,\alpha):=\E[\log \Gamma(X+\alpha)]$.
Using this notation, we may write
\begin{align} 
H(X)&=-r \log(1-p)-p r (\log p)/(1-p)+\log \Gamma(r)+\ebin(r,p,1)-\ebin(r,p,r)
\nonumber \\
&=\frac{r h(p)}{1-p}+\log \Gamma(r)+\ebin(r,p,1)-\ebin(r,p,r).
\label{eqn:HnbinDef}
\end{align}
Let $q := p/(1-p)$.
We now consider the function $Q(z)=M(\log(z+1))$, which can be written as
\begin{align}
Q(z)&=\left(\frac{1-p}{1-p (z+1)}\right)^r 
= \left(\frac{1}{1-q z}\right)^r \label{eqn:qnbin}  \\
&= \sum_{j=0}^\infty \binom{-r}{j} (-qz)^j 
= \sum_{j=0}^\infty \binom{j+r-1}{j} (qz)^j \nonumber,
\end{align}
and thus the $j$th power series coefficient of $Q(z)$; i.e., 
the $j$th factorial moment of the distribution is equal to
\[
j! \binom{j+r-1}{j} q^j.
\]
We are now ready to apply Theorem~\ref{thm:logG} and conclude,
using \eqref{eqn:ElogGa}, that
\begin{equation} \label{eqn:enbin}
\enbin(r,p,\alpha) = \log \Gamma(\alpha)+
\sum_{j=1}^\infty \binom{j+r-1}{j} (-q)^j c_\alpha(j).
\end{equation}
Plugging this result into \eqref{eqn:HnbinDef}, we
thus have verified \eqref{eqn:hnbinSummary}:
\begin{align*} 
H(X)&=\frac{r h(p)}{1-p}+
\sum_{j=1}^\infty \binom{j+r-1}{j} \left(\frac{p}{p-1} \right)^j (c(j)-c_r(j)).
\end{align*}
Notice that, when $r=1$, the above expression reduces to $H(X)=h(p)/(1-p)$, which
is the entropy of a geometric distribution.

Let us now consider the logarithmic expectation of the distribution.
Define $\enbin'(r,p,\alpha):=\E[\log(X+\alpha)]$. Similar to 
$\enbin$, we can use \eqref{eqn:ElogGd} to expand
$\enbin'$ as
\begin{align}
\enbin'(r,p,\alpha) &= 
-\sum_{j=1}^\infty \binom{j+r-2}{j-1} (-q)^{j-1} c_\alpha(j) \nonumber \\
&= 
\frac{1}{q}\sum_{j=0}^\infty \binom{j+r-2}{j-1} (-q)^{j} c_\alpha(j). 
\label{eqn:logEnbin}
\end{align}
When $r=1$, this reduces to the logarithmic expectation of a geometric
distribution. Let us write
\begin{align}
\egeom(p,\alpha)&:=\ebin'(1,p,\alpha) \nonumber\\
&\stackrel{\eqref{eqn:logEnbin}}{=}
\frac{1}{q}\sum_{j=0}^\infty (-q)^{j} c_\alpha(j) \nonumber\\
&\stackrel{\eqref{eqn:Cgen}}{=} \frac{1}{q}C_\alpha(-q) \label{eqn:geomCa}\\
&\stackrel{\eqref{eqn:CgenLerch}}{=}
-\frac{1}{1+q} \Phi'_\alpha\left(\frac{q}{1+q}\right) \nonumber \\
&=(p-1) \Phi'_\alpha(p), \label{eqn:enbinG}
\end{align}
%
where $C_\alpha(\cdot)$ is the generating function 
of the logarithmic difference coefficients defined in 
\eqref{eqn:Cgen} and $\Phi'_\alpha$ is the derivative of
the Lerch transcendent (and the polylogarithm for $\alpha=1$)
defined in \eqref{eqn:lerch} and \eqref{eqn:plog}.
We have thus shown the following characterization of the
generating function of the logarithmic difference coefficients:
\begin{coro} \label{coro:coeff}
Let $\egeom(p,\alpha):=\E[\log(X+\alpha)]$
where $X \geq 0$ is geometrically distributed with mean $q=p/(1-p)$.
Then,
\[
q \egeom(p,\alpha) = C_\alpha(-q),
\]
where $C_\alpha(\cdot)$ is the generating function of the
coefficients $c_\alpha(j)$ defined in \eqref{eqn:coeff}.
\end{coro}
By combining Corollary~\ref{coro:coeff}, \eqref{eqn:CgenLerch},
and \eqref{eqn:LaplaceP}, 
and simple manipulations,
we arrive at the
following curious result:
\begin{thm} \label{thm:LaplaceG}
Let $\egeom(p):=\E[\log(X+1)]$,
where $X \geq 0$ is geometrically distributed with mean $p/(1-p)$.
Similarly, define $\epoi(\lam)=\E[\log Y!]$,
where $Y$ is a Poisson-distributed random variable with mean $\lam$.
Then, the Laplace transform of $\epoi$ 
is given by 
\[
\cL\{\epoi\}(z) = \frac{1}{z^2}\egeom\left(\frac{1}{1+z}\right).
\]
\end{thm}
%

In order to derive the integral representation for $H(X)$ 
given in \eqref{eqn:hnbinIntSummary},
we may apply Theorem~\ref{thm:logG} (specifically, \eqref{eqn:ElogGc}) 
with the choice of $Q(z)$ in \eqref{eqn:qnbin} to write, recalling
that $\E[X]=qr=rp/(1-p)$, 
\[
\enbin(r,p,\alpha)=
\log \Gamma(\alpha)+qr \log \alpha -
\int_0^1 
\frac{(1-z)^{\alpha-1}}{z \log(1-z)}\, ((1+qz)^{-r} +qr z-1) dz, 
\]
and thus, plugging the above into \eqref{eqn:HnbinDef},
\begin{align}
H(X)&=\frac{r h(p)}{1-p}+\log \Gamma(r)-
\int_0^1 
\frac{(1+qz)^{-r} +qr z-1}{z \log(1-z)}\,  dz \nonumber \\
&-\log \Gamma(r)-qr \log r+
\int_0^1 
\frac{(1-z)^{r-1}((1+qz)^{-r} +qr z-1)}{z \log(1-z)}\,  dz \nonumber\\
&=\frac{r(h(p)-p \log r)}{1-p} +
\int_0^1 
\frac{((1-z)^{r-1}-1)((1+pz/(1-p))^{-r} +pr z/(1-p)-1)}{z \log(1-z)}\,  dz.
\nonumber
\end{align}
which recovers \eqref{eqn:hnbinIntSummary}.

\section{Entropy of the hypergeometric distribution}
\label{sec:hg}
The hypergeometric distribution is regarded as an analogue 
of the binomial distribution, with the difference that the
Bernoulli trials are performed without replacement. Namely,
the parameters $n, N, K$ are respectively regarded as 
the number of trials, population size, and number of success
states in the population. For each trial, an item is
drawn, uniformly at random, from the population,
in which initially $K$ specific items are marked as
``success states''. 
The trial results in a success if a success state is drawn,
and is otherwise a failure. After each trial, the drawn item
is discarded from the population, and the resulting random
variable counts the number of successful trials.
The probability mass function for this distribution is given
in \eqref{eqn:hg}. 
The distribution contains the binomial distribution 
(and thus, the Poisson distribution) as 
a limiting case when the ratio $K/N=p$ is a given success probability
and $N$ tends to infinity. 
While the parameters $N,K,n$ are normally
set to be integers, the distribution (over $k=0,1,\ldots$)
still normalizes to a total mass of $1$ and is thus
well defined even if the parameters may be chosen to be
non-integral, in which case the binomial coefficients should
be understood in terms of the beta function.

As in the beta-binomial distribution, the moment 
generating function for the hypergeometric distribution
involves the hypergeometric function, and is given
by the expression 
\[
M(t) = \binom{N-K}{n}\hyper(-n,-K;N-K-n+1;e^t)\Big/\binom{N}{n}.
\]
Thus in this case, the function $Q(z)=M(\log(z+1))$ is
equal to
\[
Q(z) = 1+\sum_{j=1}^\infty q(j) z^j = 
\binom{N-K}{n}\hyper(-n,-K;N-K-n+1;1+z)\Big/\binom{N}{n}.
\]
Despite the seemingly complicated expression, the coefficients
$q(j)$, which are the factorial moments of the distributions,
we known to have a simple form \cite{ref:factorial}:
\[
q(j) = \frac{\fall{K}{j} \fall{n}{j}}{\fall{N}{j}},
\]
where $\fall{a}{b}=\binom{a}{b} b!$ denotes the falling factorial.
We are now ready to apply Theorem~\ref{thm:logG} to show that,
letting $X$ be drawn from the hypergeometric
distribution with probability mass function \eqref{eqn:hg},
\begin{equation} \label{eqn:E:hg}
E_{N,K,n}:=\E[\log X!] =  
\sum_{j=2}^\infty (-1)^j q(j) c(j) =
\sum_{j=2}^\infty (-1)^j c(j) \frac{\fall{K}{j} \fall{n}{j}}{\fall{N}{j}}.
\end{equation}
The entropy of $X_{N,K,n}$ can be directly expressed from 
\eqref{eqn:hg} as
\begin{multline} \label{eqn:H:hg:raw}
H(X)=\log \binom{N}{n}-\log K!-\log(N-K)! \\
+\E[\log X! + \log(n-X)! + \log(K-X)! + \log(N-K-n+X)!].
\end{multline}
We now used the following basic symmetries in the distribution:
\[
\hg(k;N,K,n)=\hg(n-k;N,N-K,n)=\hg(K-k;N,K,N-n)=\hg(N-K-n+k;N,N-K,N-n),
\]
where the last identity is obtained by combining the first two.
This, in turn, implies that
\begin{align*}
\E[\log(n-X)!] &= E_{N,N-K,n}\\
\E[\log(K-X)!] &= E_{N,K,N-n} \\
\E[\log(N-K-n+X)!] &= E_{N,N-K,N-n}.
\end{align*}
Plugging this result back in \eqref{eqn:H:hg:raw}, combined with
\eqref{eqn:E:hg}, leads to the 
following series for the entropy of the hypergeometric 
distribution,
\begin{align*}
H(X)&=\log \binom{N}{n}-\log K!-\log(N-K)! \\&+
\sum_{j=2}^\infty \frac{(-1)^j c(j)}{\fall{N}{j}} \left(
\fall{K}{j} \fall{n}{j} +
\fall{N-K}{j} \fall{n}{j} +
\fall{K}{j} \fall{N-n}{j} +
\fall{N-K}{j} \fall{N-n}{j}
\right),
\end{align*}
proving \eqref{eqn:hhgSummary}.

\section{Discussion} \label{sec:discussion}

In this work, we studied a general method for deriving 
series expansions and integral representations for logarithmic
and log-gamma expectations of arbitrary distributions. As
a result, we obtained entropy expressions for several
fundamental distributions, including the Poisson, binomial,
beta-binomial, and negative binomial distributions. It is 
natural to ask whether the technique can be extended to
derive clean expressions for the entropy of
other distributions.
Another natural direction is whether the techniques
can be used to obtain clean, general, and high-precision estimates
of the entropy functions in terms of elementary functions.

We have also discovered connections between 
logarithmic expectations of different distributions,
and moreover, connections between them and 
generalizations of the Riemann zeta function
via the Laplace transform. An intriguing question
is whether such curious connections with functional transforms
are isolated
facts or can be further developed into a richer theory.
Finally, our work calls for a further study and better
understanding of the logarithmic
difference coefficients $c_\alpha(j)$, which are also of interest
in the analytic study of zeta functions.

\section*{Acknowledgement}
The author thanks an anonymous reviewer for comments on related works
\cite{ref:Kne98,ref:JS99,ref:CGKK13}.

\bibliographystyle{abbrv}
\bibliography{\jobname}

\appendix

\section{Connection between the $c_\alpha(j)$, the
Riemann zeta function, and Harmonic numbers}
\label{sec:zeta}

In this appendix, we observe a connection between the
logarithmic difference coefficients and the Riemann zeta
function, leading to an interesting formula for a
harmonically weighted summation of the coefficients.
 
The following convergent Newton expansion series for Hurwitz zeta function
$\zeta(s,\alpha) := \sum_{n=0}^\infty (n+\alpha)^{-s}$ 
(which reduces to the Riemann zeta function $\zeta(s)$ at $\alpha=1$)
was given by Hasse \cite{ref:Has30}: For all $\alpha>0$ and $s \neq 1$,
\begin{equation}
\zeta(s,\alpha) (s-1)=\sum_{j=1}^\infty \frac{1}{j} 
\sum_{k=0}^{j-1} \binom{j-1}{k} (-1)^k (k+\alpha)^{1-s}.
\end{equation}
Taking the derivative of the above in $s$, denoting
$\zeta'(s,\alpha):=\frac{d}{ds} \zeta(s,\alpha)$, gives
\begin{align*}
\zeta(s,\alpha)+(s-1)\zeta'(s,\alpha)&=
-\sum_{j=1}^\infty \frac{1}{j} 
\sum_{k=0}^{j-1} \binom{j-1}{k} (-1)^k (k+\alpha)^{1-s} \log(k+\alpha).
\end{align*}
Observe, from  \eqref{eqn:coeff}, that $c_\alpha(j)$ is the
inner summation on the right hand side at $s=1$.
The limit of the left hand side of the above equality at $s=1$ can be deduced
from the first two terms of the 
Laurent series expansion of the Hurwitz zeta function,
\[
\zeta(s,\alpha)=\frac{1}{s-1}-\psi(\alpha)+\sum_{n=1}^\infty a_n (s-1)^n,
\]
and is thus equal to 
$-\psi(\alpha)$, where $\psi(\alpha)=\frac{d}{d\alpha} \log \Gamma(\alpha)
=H_{\alpha-1}-\gamma$
is the digamma function, $\gamma \approx 0.57721$ is the
Euler-Mascheroni constant, and $H_k$ is the $k$th Harmonic number.
When $\alpha$ is a positive integer, we thus have
that the left identity at $s\rightarrow 1$
is equal to $-\gamma+\sum_{i=1}^{\alpha-1} \frac{1}{i}$ 
(and $-\gamma$ when $\alpha=1$). We conclude the following
identity on the logarithmic difference coefficients:
for all $\alpha>0$,
\begin{equation}
\sum_{j=1}^\infty c_\alpha(j)/j = \psi(\alpha)= \gamma-H_{\alpha-1}.
\end{equation}
Note that the left hand side can be rewritten as
$\left.\int_{0}^t \frac{C_\alpha(z)}{z}\, dz\right|_{t=1}$,
where $C_\alpha(z)$ is the generating function of the $c_\alpha(j)$
defined in \eqref{eqn:Cgen}.

\end{document}